\documentclass[authoryear,review,3p]{elsarticle}
\usepackage{graphicx}
\usepackage{bm}
\usepackage{amssymb}
\usepackage{amsmath}
\usepackage{dcolumn,multicol}
\makeatletter

\newenvironment{figurehere}
  {\def\@captype{figure}}
  {}
\makeatother
\begin{document}
\begin{frontmatter}

\title{ The Effect of Differentiation of Prey Community on Stable Coexistence in a Three-Species Food--Web Model}

\author[label1]{E. Shchekinova, M. G. J. L{\"o}der, M. Boersma, K. H. Wiltshire}

\ead{eshcheki@awi.de}
\address[label1]{Biologische Anstalt Helgoland,
Alfred-Wegener Institute for Polar and Marine Research, Kurpromenade 201,
D-27498 Helgoland
Germany}

\date{\today}
\begin{abstract}
 Food webs with intraguild predation (IGP) are widespread in natural habitats. Their adaptation and resilience behaviour is principal for understanding restructuring of ecological communities. In spite of the importance of IGP food webs their behaviour even for the simplest 3-species systems has not been fully explored.
One fundamental question is how an increase of diversity of the lowest trophic level impacts the persistence of higher trophic levels in IGP relationships. We analyze a 3-species food web model with a heterogeneous resources and IGP. The model consists of two predators directly coupled via IGP relation and indirectly via competition for resource. The resource is subdivided into distinct subpopulations. Individuals in the subpopulations are grazed at different rates by the predators. We consider two models: an IGP module with immobilization by the top predator and an IGP module with species turnover.
We examine the effect of increasing enrichment and varying immobilization (resource transfer) rate on a stable coexistence of predators and resources.
We explore how the predictions from the basic 3-species model are altered when the IGP module is extended to multiple resource subpopulations. We investigate which parameters support a robust coexistence in the IGP system. For the case of multiple subpopulations of the resource we present a numerical comparison of the percentage of food webs with stable coexistence for different dimensionalities of the resource community. At low immobilization (transfer) rates our model predicts a stable 3-species coexistence only for intermediate enrichment meanwhile at high rates a large set of stable equilibrium configurations is found for high enrichment as well.
\end{abstract}
\end{frontmatter}
\begin{multicols}{2}
\section{Introduction}\label{sec.1}

In spite of the prevalence and importance of omnivory food webs \citep{Pimm1978, Vadeboncoeur2005} in natural communities their
population dynamics to date remain poorly understood, even for only three species in the community. Even in simple systems a plethora of nonlinear effects such as flexible consumer behaviour \citep{Leibold2005}, intraspecific interactions between competing consumers and resources \citep{Holt1994}, inhomogeneity of the environment \citep{amarasekare2007, Janssen2007} and adaptive foraging \citep{Krivan1996,Krivan2005} precludes easy theoretical treatment
and interpretation.

One example of a non-trivial omnivory food web is a system with intraguild predation \citep{Polis1989, Finke2002, Borer2003}. Intraguild predation assumes that the same organism is both competitor and predator to another member of the food web. The IGP models encompass a rich dynamical behaviour including coexistence \citep{polis1992} and alternative stable states \citep{Holt2007, Daugherty2007}. Simple mathematical models \citep{polis1992,diehl2000,Namba2008} have been evoked in attempt to explain the persistence of IGP interactions in natural habitats. However predictions from the mathematical theory of 3-species IGP systems state that a high resource carrying capacity promotes the exclusion of intermediate trophic levels and thus destabilizes interactions \citep{diehl2001}. What is puzzling that various empirical studies of omnivory document however coexistence, but not exclusion, over the entire range of natural resource productivities \citep{Mylius2001,Borer2003}. On the basis of experimental observations a theoretical 3-species omnivory model \citep{Stoecker1985,Holt1997,diehl2001} predicts the coexistence only at superior competitive abilities of the IG prey for the communal resource \citep{diehl2001}. Yet empirical data suggest a robust persistence of IGP systems in both terrestrial \citep{Brodeur2000,Arim2004} and aquatic communities \citep{Polis1989,Mylius2001,Borer2003,Denno2003}.

Theoretical models that are focused on the aspects of stability and coexistence of species in 3-level systems with the IGP \citep{polis1992,Holt1997, Abrams1994, AbramsRoth1994, Abrams2010}, as a rule, largely reduce the complexity of interactions observed in realistic systems \citep{ Thomson2007}. Such oversimplifications can influence the population dynamics as well as critically impact species persistence.  Even though the simplest model of the IGP encompasses only three species \citep{polis1992,Holt1997,diehl2000,diehl2001} a number of empirical studies deal with larger food webs that involve more than three species potentially engaged in IGP interactions \citep{rosenheim1993,woodward2005}.

Spatiotemporal heterogeneity of the environment often is invoked as one of the explanatory mechanisms for the coexistence between multiple species competing for the same resources \citep{Hutchinson1961}. It has been observed that such a spatiotemporal heterogeneity can affect the diversity in prey populations \citep{amarasekare2006}. Indeed an inhomogeneity in prey items that share common resource and predators is critical in determining the responses of ecological community. For systems with multiple prey composition various coexistence patterns can be found depending on the levels of resource productivity~\citep{Leibold1996}. It is not clear yet
 how the diversity in a prey community will affect the behaviour in the IGP systems.

The effect of a habitat structure on the IGP is discussed in various recent models \citep{amarasekare2006, amarasekare2007,Janssen2007}. For example a  stable coexistence of the intraguild prey due to inhomogeneity of a habitat can be supported by creating temporal refuges for prey and reducing the encounter rates among preys and predators \citep{Janssen2007}. In addition the stability of the IGP can be enhanced by an inclusion of additional factors such as behaviourally mediated effects \citep{Janssen2007}.

To include the effect of an increasing diversity of resource and IG predators on population dynamics recently the 3-species IGP model \citep{Holt1997} was modified by Holt and Huxel (2007). The authors extended the basic 3-species omnivory model to the so called "partial IGP" model in which "partial" overlap among competitors for a single resource exists and both predators have exclusive resources to exploit. It was shown \citep{Holt2007} that an alternative resource enhances the tolerance of the IG prey against attacks from IG predators. Independently of a competitive status of the IG prey in exploitation for a shared resource it can persists by utilizing an alternative resource.
An extended formulation of the IGP model with trophic supplementation has been proposed by Daugherty et al. (2007). The authors investigated three forms of a supplementary feeding outside of the basic IGP module and postulated
a higher potential for persistence of the IG prey due to its efficient exploitation of external resources.

There is growing evidence that in many systems the IG prey has a mutualistic or at least facilitative relationship with the IG predator \citep{Crowley2011}. Including such facilitation in ecological theory will fundamentally change many basic predictions and will enable a better understanding of functioning of many natural communities \citep{Bruno2003}.

Especially in the IGP systems an emphasis should be given to the elucidation of the effects of facilitation on community composition and stability \citep{Crowley2011}. Contrary to the competitive exclusion principle in systems with competitors for a single resource stability   stems from commensalism \citep{Hosack2009}. Hereby one consumer can in some way alter the habitat to benefit the other. Recently such an interaction was observed in experiments with a microzooplankton food web community \citep{Loeder2012}. The experimental system included two predators: a tintinnid species {\it Favella ehrenbergii} and a heterotrophic dinoflagellate species {\it Gyrodinium dominans}. They are both grazing on a phototrophic dinoflagellate {\it Scrippsiella trochoidea}. The authors showed that the IG predator {\it F. ehrenbergii} can precondition a substantial part of the common resource {\it S. trochoidea} during its feeding procedure by immobilizing the common prey without ingestion. Such preconditioned individuals can be captured more easily by the IG prey {\it G. dominans} than the mobile individuals of the same resource species. This mutualistic interaction leads to higher growth rates of the IG prey in the presence of the IG predator. The authors characterized their experimental observations as a facilitative IGP relationship with a commensalistic pattern.
 Our motivation for this modeling study was to investigate if such commensalistic patterns can create loopholes for a stable coexistence of all species in the investigated system. Of our major interest was if in the IGP system an immobilization \citep{Loeder2012} or the partitioning of prey populations into distinct groups of individuals offers opportunities for competition avoidance among both consumer species.

We reformulated the 3-species IGP model proposed in \citep{polis1992,Holt1997,diehl2000,diehl2001} to include multiple subpopulations of prey.
Furthermore, we explored the effect of diversification of the resource available to higher level consumers on the species persistence by numerical simulations of an extended IGP module. Specifically, we investigate how the addition of new links to a focal IGP module enhances stability of population dynamics by reducing the competitive interactions of predators for their shared resource.

In order to explain the results of the experimental findings of L{\"o}eder et al. we investigated the influence of multiple traits of the resource community on a stable coexistence in the 3-species model with different types of resource. For this purpose we adapted and reformulated the original model by Holt and Polis (1997) and added a new type of interaction. This link specifies the immobilization mechanism that depends on the densities of mobile and immobile resource items and the top predator which creates the immobile resource fraction during feeding. The immobilization term is used to model the interactions between the IG predator and the resource.

Another type of interaction considered in this paper is a resource turnover mechanism. This mechanism describes mutual interactions between species from distinct resource subpopulations. The interaction term depends exclusively on the resource subpopulation densities. The rate of turnover is constant. If no turnover or immobilization of individuals from one group to another occurs then the basic IGP model with a single population of resource is recovered. We discuss the influence of immobilization and transfer of species on the coexistence patterns in a system with different subpopulations of the resource and compare the results with the basic 3-species IGP.

This paper is organized as follows: in the first section we introduce a general 3-species IGP model with a new type of interaction that links the resource pools to the top consumers. In the following sections two distinct IGP formulations with $n=2$ resource subpopulations are discussed. Both models are derived from the basic IGP module by including additional links: $(i)$ the immobilization by the predator and $(ii)$ the  resource turnover. In the Results section we numerically investigate stability of equilibrium densities for various trophic configurations. Data from numerical analysis are presented for the IGP model with the immobilization and for the model with the resource turnover. At last we discuss results for a general IGP model with the resource turnover mechanism and $n>2$ subpopulations of the resource. After sketching the main conclusions we review the model predictions and compare their relevance to the immobilization experiment \citep{Loeder2012}. Furthermore we discuss possible alternative reformulations of the model. In the Appendix explicit forms for the steady states for two simple analytical cases and multidimensional system are specified. As a part of a linear stability routine the Jacobian matrices for two types of formulations are given. Finally, we carry over to a higher dimensional formulation and describe the parameters choice and the equilibrium densities.

\section{General model}\label{sec.2}

We introduce an omnivory model with an IGP unit derived from a simple non-spatial Lotka-Volterra system with the linear functional responses adapted from Holt and Polis (1997). The original model consists of populations of two predators (IG predator and IG prey) and a common resource. Here, we include new features such as a resource differentiation mechanism which affects palatability of a fraction of resource for the predators. Specifically, the entire resource population is subdivided into distinct groups under the assumption that the groups differ from each other by the quality and fitness of the individuals. They are consumed by the predators at different group-specific grazing rates. The differentiation of the resource could be due to damage by the predator or initial inhomogeneous distribution of the resource quality. Afterwards, we generalize our model to the case of the multiple resources.

The food web model for a multiple number of prey subpopulations $\{S_k\}^n_{k=1}$ is sketched in Fig.~\ref{fig1}~a.\footnote {Here and everywhere in the text the numerical subscripts denote species at the same trophic level.} The top predator $F$ and the intermediate predator $G$ are engaged in the IGP and share a common resource $S_1$. The resource pools are not independent because there is an exchange of individuals among different subpopulations $\{S_k\}^n_{k=2}$ following the links in Fig.~\ref{fig1}a. Another special case of the IGP with two distinct populations of resource $S_m$ and $S_i$ is presented in Fig.~\ref{fig1}b. Shown is a schematic view of trophic interactions including intraguild predation and two populations $S_i$ and $S_m$ of immobilized and mobile resources respectively. The IG prey $G$ competes with the IG predator $F$ for both resource types and is also an additional resource for the IG predator. The size of the population $S_i$ increases due to immobilization of individuals from the population $S_m$ by the IG predator $F$.

We begin with an overview of a general IGP model and all the important trophic links and parameters that are used to define it. Later we focus specifically on two different formulations of the general IGP model.

The general model for a food web with an inhomogeneous resource is derived from the Lotka-Volterra omnivory model \citep{diehl2000,diehl2001} with the interaction term
that accounts for the transitions among different pools. The Lotka-Volterra omnivory model consists of $n+2$ equations. It is used as an approximation for the food web community with the IGP and $n\ge2$ mutually interacting subpopulations of the resources. In the absence of predation a basal population $S_1$ develops according to logistic growth \citep{diehl2000}. The set of equations for the population densities are written as follows:

{ \it Shared $1$st resource }:
\begin{eqnarray}
\frac{dS_1}{dt}&=&[r(1-S_1 K^{-1})-a G -f F] S_1\nonumber\\& &-z_1(S_1,S_2,\ldots,S_n,G,F),\nonumber
\end{eqnarray}
{\it Shared $k$th resource ($k=2 \ldots n$) }:
\begin{eqnarray}
\frac{dS_k}{dt}&=&w_k(S_1,S_2\ldots,S_n,G,F)\nonumber\\&&-[b_k G+f_k F+m_k] S_k,\nonumber
\end{eqnarray}
{\it Intermediate predator (IG prey)}:
\begin{eqnarray}
\frac{dG}{dt}&=&z_2(S_1,S_2,\ldots,S_n,G,F)\nonumber\\&&-(g F+m_g )G ,\nonumber
\end{eqnarray}
{ \it Top predator (IG predator)}:
\begin{eqnarray}
\frac{dF}{dt}&=&z_3(S_1,S_2,\ldots,S_n,G,F)\nonumber\\&&+(g'g G-m_f) F,\label{eq.1}
\end{eqnarray}

The parameters of the model and main populations are described in details in Table~\ref{tab1}.
Here $r$ is the maximum specific growth rate of the resource population $S_1$, $K$ is the carrying capacity of the resource defined as enrichment factor in the previous models \citep{diehl2000,diehl2001}.
The subpopulations $\{S_k\}^n_{k=2}$ are derived from
the basal resource $S_1$ via immobilization or via  individual-to-individual turnover. Species from $S_1$ and $\{S_k\}_{k\neq 1}$ are consumed by the IG predator at potentially different rates $f$ and $\{f_k\}^n_{k=2}$ and by the IG prey at rates $a$ and $\{b_k\}^n_{k=2}$ respectively. The differentiation among subpopulations $\{S_k\}^n_{k=2}$ is preserved by a choice of distinct predation pressures $\{b_k\}^n_{k=2},\{f_k\}^n_{k=2}$, feeding rates $\{f'_k\}^{n}_{k=2}, \{b'_k\}^{n}_{k=2}$ and mortality coefficients $\{m_k\}^{n}_{k=2}$. The density-independent mortality rates for $S_1, G$ and $F$ are $m_1,m_g$ and $m_f$ correspondingly. They are used as factors limiting the growth of the populations in (\ref{eq.1}).

A key assumption of the model is that there is only one-directional movement between the basal resource $S_1$ and its fractions $\{S_k\}_{k\neq 1}$.  The local interactions among individuals from alternative pools are embedded via functional terms $\{w_k\}^n_{k=2}$ provided in Table~\ref{tab2} for each type of the IGP formulation. These terms account for
transitions among the resource items $\{S_k\}_{k\neq 1}$. The general omnivory model (\ref{eq.1}) can be reduced to three types of IGP formulations: system with immobilization and systems with the resource turnover for $n=2$ subpopulations and for $n>2$ pools. For each of the formulations specific expressions of functional forms $z_1, z_2, z_3$ and $\{w_k\}$
 are provided in the Table~\ref{tab2}. The term $z_1$ is responsible for the exchange of individuals among subpopulations $\{S_k\}$ due to the species turnover or the immobilization mechanism. The transfer of individuals from the population $S_1$ to $\{S_k\}^n_{k=2}$ happens instantaneously  at constant rates $\{c_k\}^n_{k=2}$ correspondingly. Analogously $\{q_{k,j}\}_{k\neq j}$ are defined as instantaneous migration rates among subpopulations $\{S_k\}^n_{k=2}$. The terms $z_2$ and $z_3$ are used to evaluate the total predation of the IG prey and the IG predator on the resource.

To achieve a stable persistence of all species the IG prey should benefit more from an alternative resource than the IG predator. For this reason, whereas the attack rates of the IG predator are equal for different resource pools, the IG prey establishes a higher predation pressure on subpopulations $\{S_k\}_{k\neq 1}$ than on the basal pool $S_1$. The numerical values for the attack rates are chosen to be close to the experimentally observed values \citep{Loeder2012}.

Holt and Huxel (2007) used an extended IGP module with alternative resources that are defined independently. They evolve according to their own intrinsic growth rates. As opposed to the formulation given by Holt and Huxel (2007) and to a model with trophic supplements \citep{Daugherty2007} here we do not consider external alternative resources. In our model with immobilization the population density in every resource pool varies due to immobilization by the IG predator and consumption by the predators. Similarly in the formulation with the resource turnover the transfer mechanism between resource subpopulations plays a role in exchange among the distinct resource pools. Alternative pools grow due to the influx of species from the basal resource or the other pools. Therefore the sizes of subpopulations are controlled mainly by a number of direct encounters with the IG predator (immobilization) or by a species turnover from one resource subpopulation to another. In addition, the individuals in the different pools of the basal resource are distinguished by group-specific predation pressures that establish a top-down regulation of densities of each subpopulation.

In the following sections we present an explicit formulation of the model with immobilization and of the model with the resource turnover for $n=2$ subpopulations.

\begin{figurehere}
\begin{center}
  \includegraphics[width=8cm]{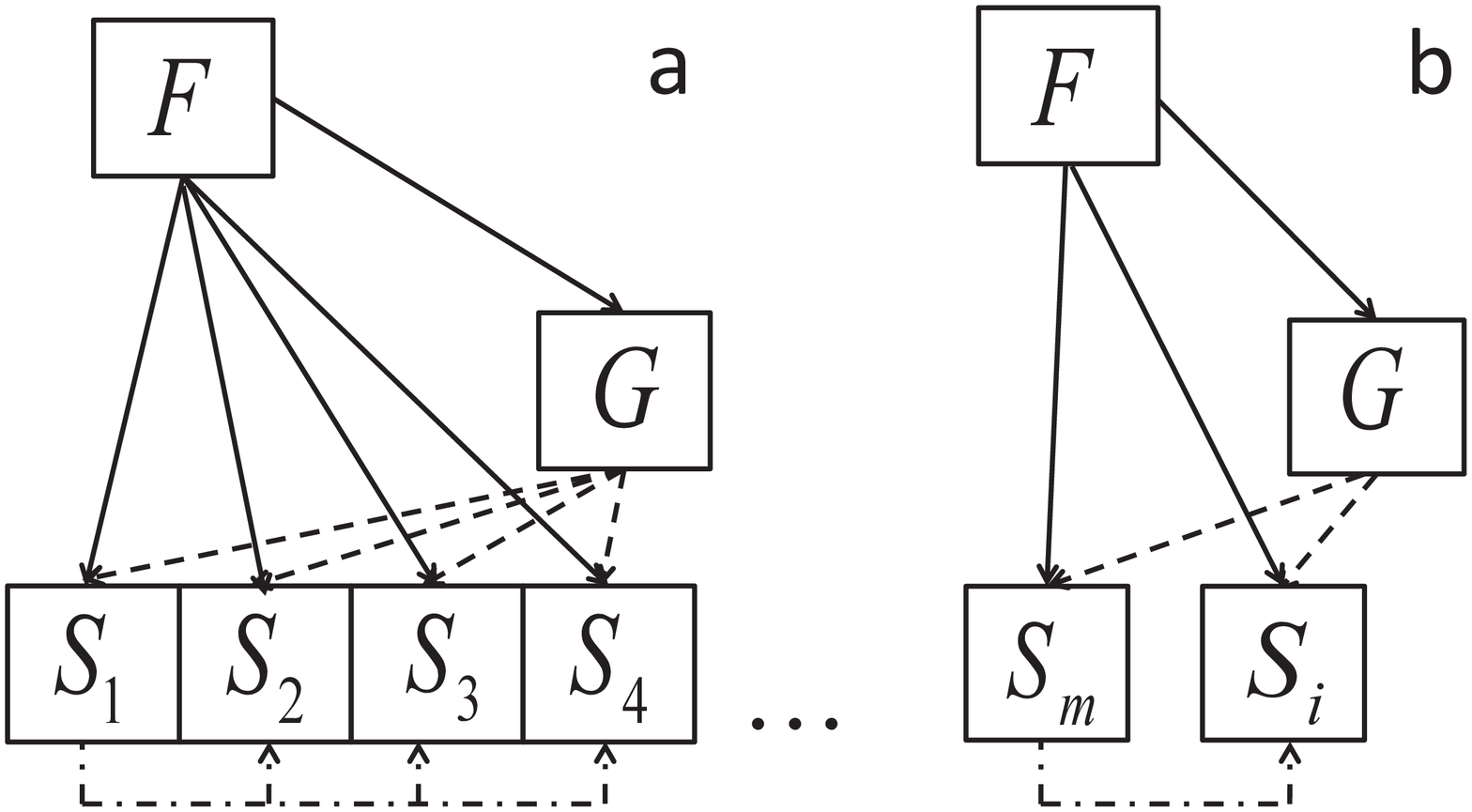}
  \end{center}
  \caption{(\textbf{a}) General structure of the food web model with two predators and multiple resource subpopulations; (\textbf{b}) the structure of the food web with the presence of the immobilization mechanism by the IG predator. The resource is subdivided into populations of mobile $S_m$ and immobilized $S_i$ individuals. The links represent: (solid) food resources for the top predator, (dashed) food resources for the intermediate predator, (dot-dashed) transitions between different resource pools.}\label{fig1}
\end{figurehere}

\end{multicols}

 \vspace{.5cm}
 \par\noindent\rule{\dimexpr(0.5\textwidth-0.5\columnsep-0.4pt)}{0.4pt}%
    \rule{0.4pt}{6pt}
 \begin{table}
\caption{\label{tab1} The variables and parameters for the general model ~(\ref{eq.1}).} \vspace{0.2cm}
\begin{tabular}{l l }
\hline
  $ $&$ $\\
    & Definition\\
  $ $&$ $\\
  \hline
  &General model\\
$Populations $&$ $\\
  $S_k$ & population size of resource in the $k$th pool,\\
  $ G$& population size of IG prey \\
  $ F$& population size of IG predator\\
  $Parameters$&\\
  $r$ & maximum specific growth rate of the resource $S_1$ \\
  $K$& resource carrying capacity or enrichment\\
  $a$& attack rate of predator $G$ on $S_1$ subpopulation \\
  $f$& attack rate of predator $F$ on $S_1$   \\
  $g$ & attack rate of predator $F$ on $G$ \\
$c_k$& per capita effect of species $S_1$ on $S_k$ \\

$q_{k,j}$ & per capita effect of species $S_k$ to $S_j$ \\
$b_k$ & attack rate of predator $G$ on $S_k$ species \\
$f_k$ & attack rate of predator $F$ on $S_k$ species \\
$m_k$ &    mortality rate of species from $S_k$ subpopulation \\
$m_g$ & mortality rate of $G$ \\
$m_f$ & mortality rate of $F$ \\
$g'$ & converting efficiency of food resource $G$ into $F$ \\
$f'_j$ & growth rate of $F$ from resource $S_j$  \\
$b'_j$ & growth rate of $G$ from resource $S_j$  \\
$ $  & $$ \\
  \hline
\end{tabular}
\end{table}
\hfill
\vspace{\belowdisplayskip}\hfill\rule[-6pt]{0.4pt}{6.4pt}%
    \rule{\dimexpr(0.5\textwidth-0.5\columnsep-1pt)}{0.4pt}
\vspace{.5cm}

 \par\noindent\rule{\dimexpr(0.5\textwidth-0.5\columnsep-0.4pt)}{0.4pt}%
    \rule{0.4pt}{6pt}
 \begin{table}
\caption{\label{tab2} The variables and parameters for the models ~(\ref{eq.2}) and (\ref{eq.3}).} \vspace{0.2cm}
\begin{tabular}{l l }
\hline
  $ $&$ $\\
    & Definition\\
  $ $&$ $\\
  \hline&Model with immobilization\\
  $ Populations $&$ $\\
$S_m, S_i$ & population sizes of mobile (immobilized) resource\\

  $Parameters$&\\
  $r$ & maximum specific growth rate of population $S_m$ \\
$K$ & resource carrying capacity \\
$a,b$ & attack rates of predator $G$ on mobile (immobilized) population \\
$f$ & attack rate of predator $F$ on mobile and immobilized populations \\
$i_m$ &immobilization rate \\

$a', f'$ & conversion efficiency factors \\
$ $  & $$ \\
&Model with a resource turnover \\
  $ Populations $&$ $\\
$S_1, S_2$ & population sizes of resources\\

$Parameters$&\\
 $r$ & maximum specific growth rate of subpopulation $S_1$ \\
$K$ & resource carrying capacity \\
$a, b$ & attack rates of predator $G$ on subpopulations $S_1, S_2$\\
$f$ & attack rate of predator $F$ on subpopulations $S_1$ and $S_2$\\
$t_r$ & transfer rate or per capita effect of $S_1$ on $S_2$\\

$ $& $$ \\

  \hline
\end{tabular}
\end{table}
\hfill
\vspace{\belowdisplayskip}\hfill\rule[-6pt]{0.4pt}{6.4pt}%
    \rule{\dimexpr(0.5\textwidth-0.5\columnsep-1pt)}{0.4pt}
\begin{multicols}{2}
\end{multicols}

 \vspace{.5cm}
 \par\noindent\rule{\dimexpr(0.5\textwidth-0.5\columnsep-0.4pt)}{0.4pt}%
    \rule{0.4pt}{6pt}
 \begin{table}
\caption{\label{tab3} Description of the functional forms used in the system~(\ref{eq.1}).} \vspace{0.2cm}
\begin{tabular}{l l }
\hline
  $ $&$ $\\
  Description & Model equations\\
  $ $&$ $\\
  \hline
  $ $&$ $\\
  System with immobilization: & $S_1=S_m, S_2=S_i,z_1=i_m S_m F,~w_2=z_1, z_2=a'(a S_m+b S_i)G$,\\
  $ $&$~z_3=f'f(S_m+S_i)F$ \\
  $ $&$ $\\
  System with resource turnover:& $z_1=t_r S_1 S_2,~w_2=z_1,~z_2=a'(a S_1+b S_2) G,~z_3=f'f(S_1+S_2)F$\\($n=2$ resource subpopulations) &$ $ \\
  $ $&$ $\\
  System with the resource turnover: & $z_1=S_1 \sum_{k=2}^{n} c_k S_k,~z_2= G\sum_{j=1}^{n} b'_j S_j, z_3=F\sum_{j=1}^{n} f'_j S_j $,\\($ n>2$ resource subpopulations) &$w_k=S_k(c_k S_1+\sum_{j=2}^{n} q_{k,j} S_j), ~k=2\ldots n$ \\
  $ $ & $ $\\
  $ $&$ $\\
  \hline
\end{tabular}
\end{table}
\hfill
\vspace{\belowdisplayskip}\hfill\rule[-6pt]{0.4pt}{6.4pt}%
    \rule{\dimexpr(0.5\textwidth-0.5\columnsep-1pt)}{0.4pt}
\begin{multicols}{2}
\subsection{System with immobilization by predator}

The system with the immobilization illustrated in Fig.~\ref{fig1}~b is derived from the equations (\ref{eq.1}) for two resource subpopulations by substituting the interaction terms $z_1,z_2,z_3$ and $w_2$ from Table~\ref{tab1}.
After the substitution the set of equations for the IGP model with immobilization yields:\\
{\it Mobile resource:}
\begin{eqnarray}
\frac{dS_m}{dt}&=&[r(1-S_m K^{-1})-a G]S_m\nonumber\\&&-(f+i_m) F S_m,\nonumber
\end{eqnarray}
{\it Immobilized resource:}
\begin{eqnarray}
\frac{dS_i}{dt}&=&i_m F S_m-[b G+f F] S_i ,\nonumber
\end{eqnarray}
{\it IG prey:}
\begin{eqnarray}
\frac{dG}{dt}&=&[ a'a S_m +a'b S_i -g F-m_g ] G,\nonumber
\end{eqnarray}
{\it IG predator:}
\begin{eqnarray}
\frac{dF}{dt}&=&[f'f(S_i+S_m)+g'g G -m_f] F,\label{eq.2}
\end{eqnarray}
where the state variables $S_m$ and $S_i$ are the densities of mobile and immobilized species.
Note that the feeding rates of the top predator $F$
on both populations $S_i$ and $S_m$ are equal. By contrast, the attack rate of the IG prey on immobilized subpopulation is higher than on mobile species. The relation $b>a$ holds in the presence and in the absence of the predator $F$.
This assumption is well justified by the observations of an experiment with artificial immobilization \citep{Loeder2012}. {\it G. dominans} demonstrate a strongly selective behaviour towards immobilized species when offered in a mixture with mobile cells of {\it S. trochoidea}. It was measured that ingestion rates of the predator in the immobilized prey treatment were by a factor of $20$ greater than those in the control treatment.

The stability of equilibrium densities and the persistence zones of the system~(\ref{eq.2}) with a non-zero immobilization rate are discussed in Section~\ref{sec.3.1}.

\subsection{System with the resource turnover}
\subsubsection{General case of $n=2$ subpopulations}
 The model with the resource turnover is derived from the general case~(\ref{eq.1}) by substituting the functional forms from Table~\ref{tab2}. It is written as follows:\\
{\it $1$st resource}:
\begin{eqnarray}
\frac{dS_1}{dt}&=&[r(1-S_1 K^{-1})-a G] S_1\nonumber\\
&& -[f F+t_r S_2] S_1\nonumber,
\end{eqnarray}
{\it $2$nd resource}:
\begin{eqnarray}
\frac{dS_2}{dt}&=&[t_r S_1 -b G-f F] S_2,\nonumber
\end{eqnarray}
{\it IG prey}:
\begin{eqnarray}
\frac{dG}{dt}&=&[a'a S_1+a'b S_2-g F-m_g ] G,\nonumber
\end{eqnarray}
{\it IG predator}:
\begin{eqnarray}
\frac{dF}{dt}&=&[f'f (S_1+S_2)+g'g G -m_f] F.\label{eq.3}
\end{eqnarray}
All the parameters are chosen the same as for the system with immobilization~(\ref{eq.2}). Note that the evolution equations are written as in ~(\ref{eq.2}) but immobilization term is replaced with the transfer term that is dependent on the population densities. The transfer between the two subpopulations occurs each time whenever species from two different pools encounter each other. In the simplest case the number of encounters is proportional to the population densities of $S_1$ and $S_2$.

 If the density of second subpopulation is zero and no differentiation in the resource takes place at $t_r=0$ than the top predator $F$ outcompetes the predator $G$ due to a higher predation rate ($f>a$). This outcome is predicted by the basic IGP model \citep{ diehl2000,diehl2001}. By contrast, whenever the turnover of species takes place and non-zero densities are produced in the resource pool $S_2$ the intraguild predation introduces a higher pressure on the second subpopulation $S_2$. This will potentially lead to a negative effect on the population density in $S_2$ and to higher levels of subpopulation $S_1$. The result of this interaction is that the 3-species coexistence is reached via the IGP competition trade-off.

\section{Main results}\label{sec.3}

We illustrate an emergent dynamical behaviour for the three formulations provided in Table~\ref{tab3} with stability diagrams. Due to high dimensionality of the models (\ref{eq.1})-(\ref{eq.3}) the analysis of an entire parameter space is intractable. Only several illustrative examples for every formulation will be shown here.

\subsection{Model with immobilization}\label{sec.3.1}

In Fig.~\ref{fig2} the regions of stable positive equilibrium solutions versus immobilization and enrichment are shown. The parameter space is partitioned into several stability zones associated with the regions of coexistence, exclusion of both predators and exclusion of the IG prey at $G=0$. The boundaries defined for partitioning of the diagram are found from the eigenvalue analysis (see Appendix). As shown in  Fig.~\ref{fig2} at low enrichment the densities of both predators decay to zero and  the summed abundance of the resource reaches steady state at $S_m+S_i=K$. The case of zero immobilization has been already considered in previous studies \citep{Holt1997,diehl2000}. At low immobilization and at high enrichment only the top predator and resource are stable and positive, just as in the 3-species IGP model \citep{diehl2000}, whereas the coexistence between both predators and common resource is possible only in the regions of intermediate enrichment. A higher mortality rate for the predator $G$ results in its extinction in the region of low immobilization in Fig.~\ref{fig2}b where an extra resource can no longer support its persistence. Only the IG predator and resource persist in this region of parameters. Situation is different for higher immobilization where a large region of coexistence for both predators exists. The equilibrium densities shown on the diagrams are defined in the eq.~(\ref{eq.7}) in Appendix.

 Fig.~\ref{fig3} shows equilibrium densities of the four components of the food web and their dependence on enrichment and immobilization rates.  For a high immobilization rate the resource population is dominated by immobilized individuals. Meanwhile at low immobilization mobile and immobilized populations increase along the gradient of enrichment an adverse pattern occurs at high immobilization. The growth rate of the IG predator is noticeably reduced at $i_m=0.2$ due to an increase of the competitive trade-off with the IG prey.
The dependence of the population densities on the enrichment of resource is shown in Fig.~\ref{fig4} for the immobilization $i_m=0.3$. Meanwhile as predicted from the standard IGP model \citep{diehl2000} the IG prey is excluded at high enrichment in the model with immobilization at $i_m=0.3$ the IG prey benefits from immobilized resource and its persistence is increased at a broader range of carrying capacities. The density of the mobile (immobilized) resource subpopulation reach saturation threshold at a higher enrichment (see Fig.~\ref{fig4}).
\begin{figurehere}
\begin{center}
\includegraphics[width=8.cm]{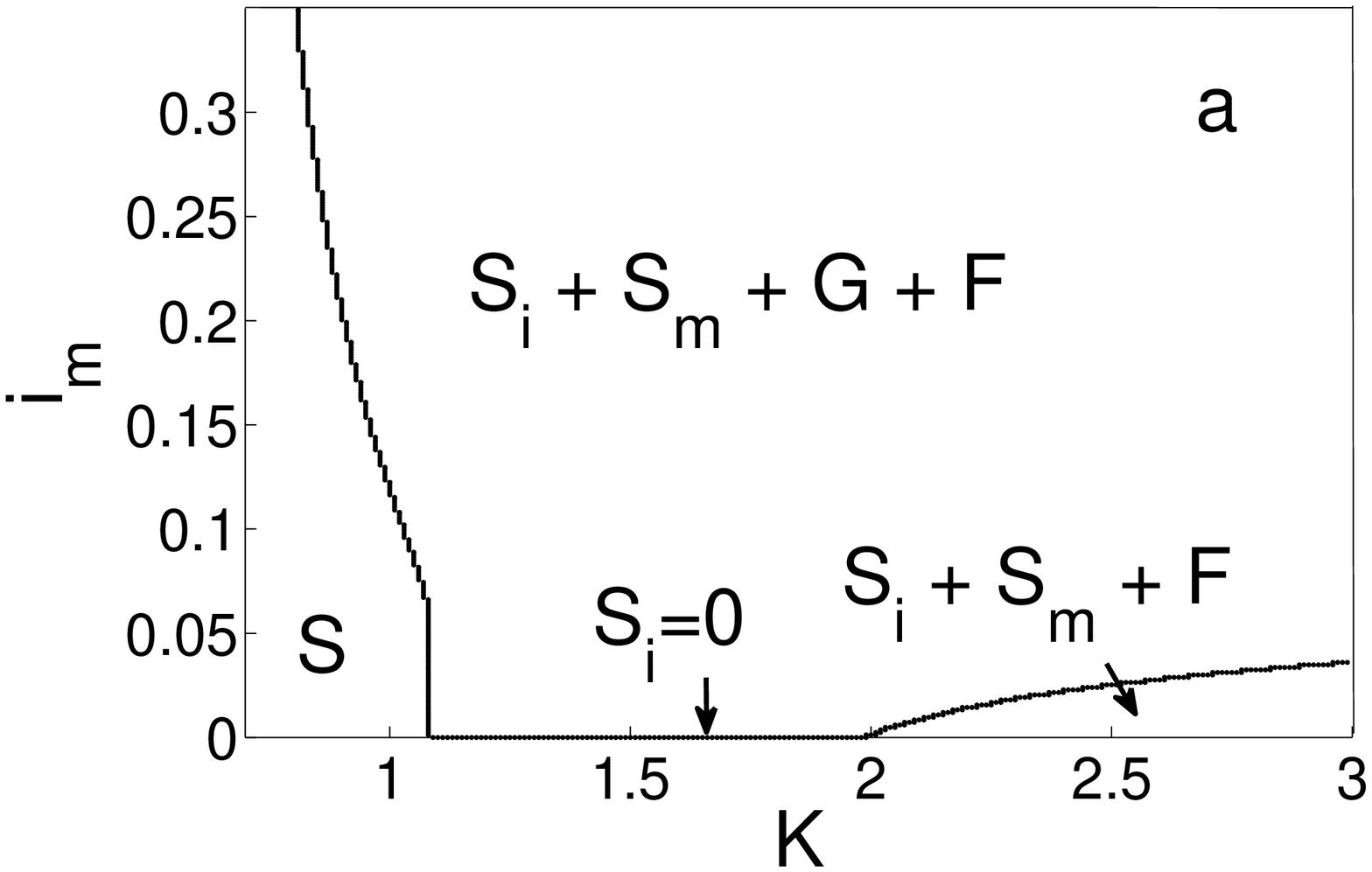}
\includegraphics[width=8.cm]{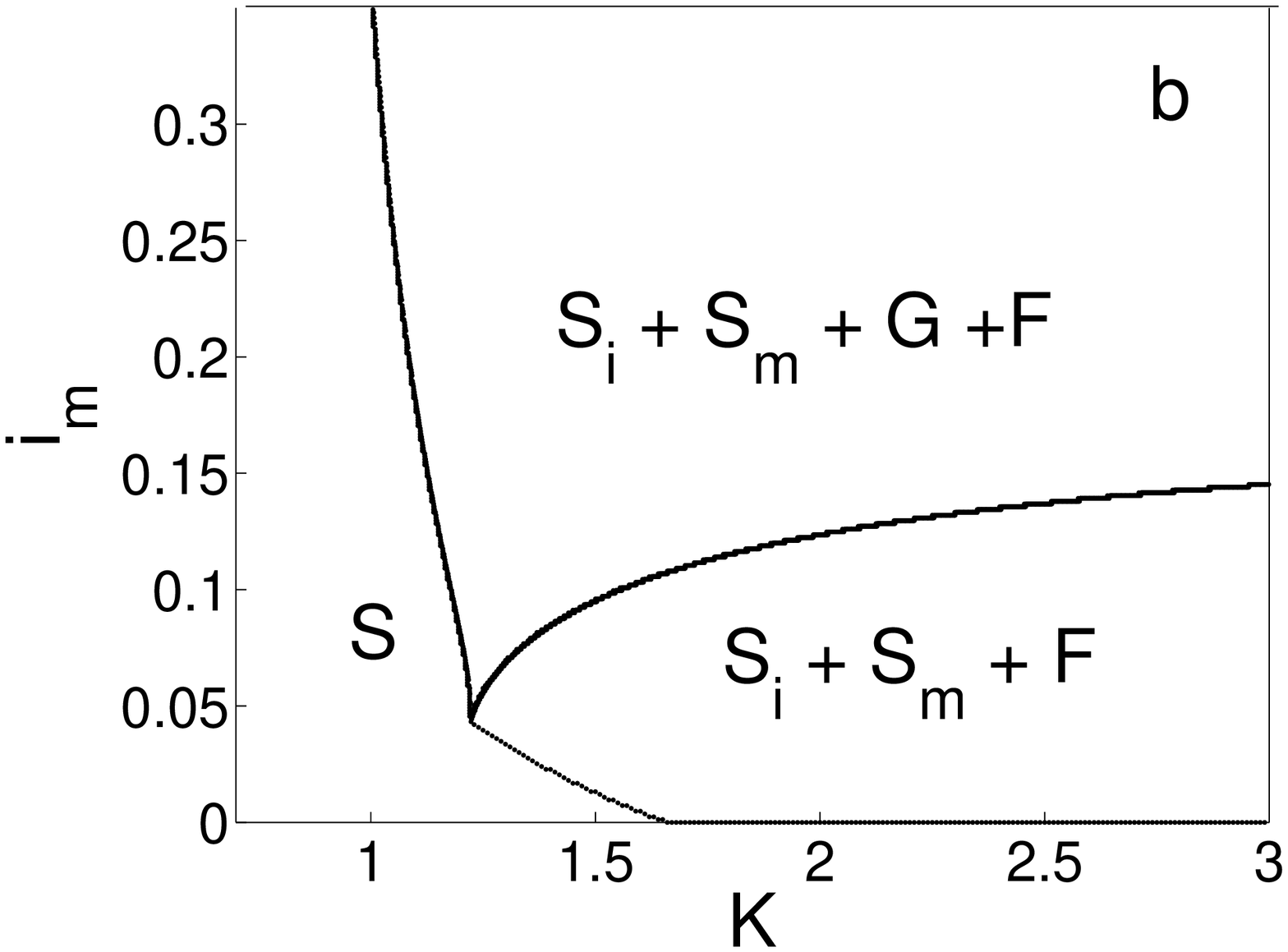}
\end{center}
  \caption{Regions of stable coexistence for the immobilization model~(\ref{eq.2}). Stability diagrams are partitioned into the regions of stable (unstable) coexistence and alternative states with exclusion of one of the predators. Letters stand for persistence of different trophic configurations: $(S)=$exclusion of both predators, $(S_i+S_m+F)=$coexistence of top predator and resource, $(S_i+S_m+F+G)=$3-species coexistence. Parameters are: $r=0.4, f=0.12, a=0.025, b=0.1, a'=0.8, g=0.025, f'=0.2, g'=0.5, m_f=0.04$. Two plots for:(\textbf{a}) $m_g=0.02$, (\textbf{b}) $m_g=0.06$ are shown.}\label{fig2}
  \end{figurehere}
\begin{figurehere}
\begin{center}
  \includegraphics[width=9.cm]{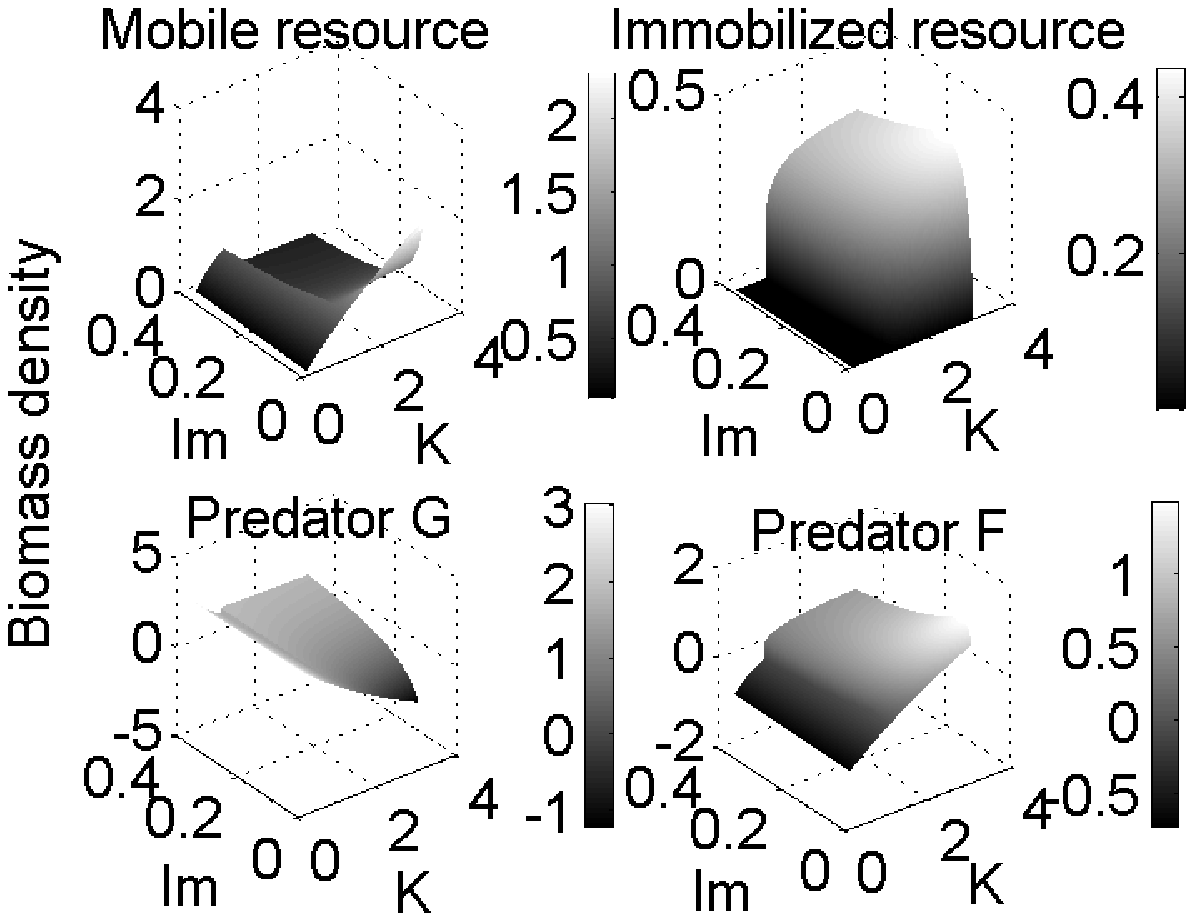}
  \end{center}
  \caption{Equilibrium biomass densities along immobilization and enrichment gradients. Shown are densities of: (\textbf{a}) mobile resource, (\textbf{b}) immobilized resource,(\textbf{c}) predator $G$,(\textbf{d}) predator $F$. Parameters are used as in Fig.~\ref{fig2}}\label{fig3}
\end{figurehere}
\begin{figurehere}
\begin{center}
\includegraphics[width=9.cm]{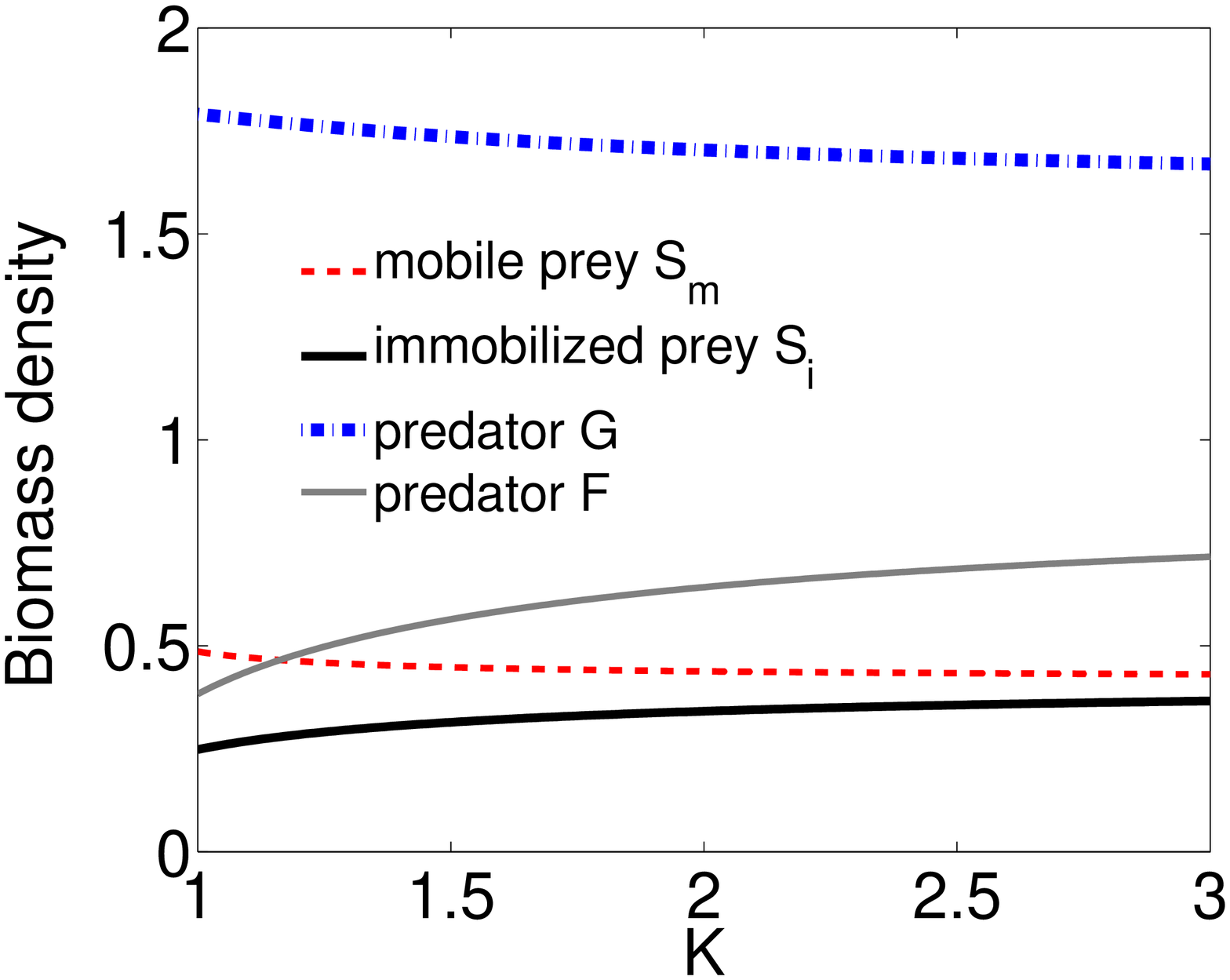}
  \end{center}
  \caption{Equilibrium biomass densities versus enrichment for fixed immobilization rate $i_m=0.3$.}\label{fig4}
\end{figurehere}

\begin{figurehere}
\begin{center}
  \includegraphics[width=9.cm]{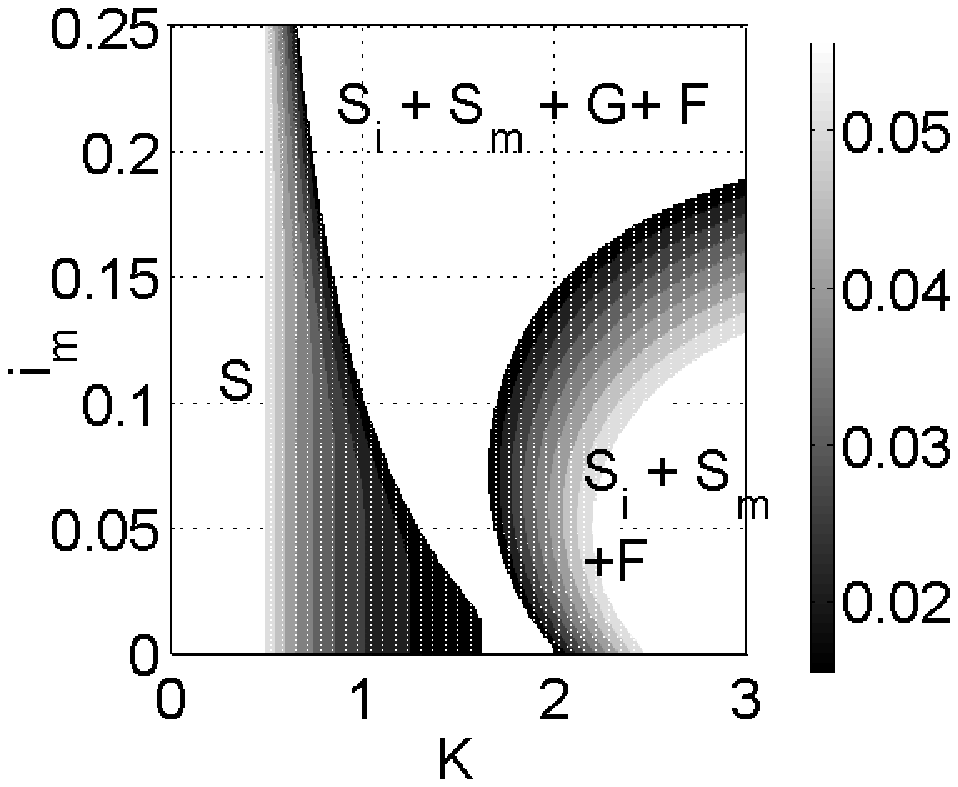}
  \end{center}
  \caption{Regions of stable coexistence for the immobilization model (\ref{eq.2}). The level curves for the grazing rate $a$ define the boundaries of the stability regions. The colorbar shows the values of $a$. Parameters are: $r=0.5, f=0.1,b=0.1,a'=0.8,g=0.07,f'=0.2,g'=0.5,m_g=0.02,m_f=0.04$}\label{fig5}
\end{figurehere}
How sensitive is a stable coexistence to small variations of the attack rates of the intermediate predator $G$? Will our predictions be still valid? To examine the system behaviour for different attack rates of $G$ the coexistence zones are exemplified for different values of $a$ in Fig.~\ref{fig5}. Colorcode is assigned according to grazing rate $a$. Overall the stability diagram exhibits similar pattern as in Fig.~\ref{fig2}. Specifically, the region of stable coexistence enlarges for higher immobilization. As it seems reasonable the number of stable solutions and the 3-species permanence zone in Fig.~\ref{fig5} gradually broadens with the increase of predation pressure from predator $G$. Simultaneously fewer exclusion steady states
for the predator $G$ are discovered.
\subsection{Model with a resource turnover}
\subsubsection{Case of $n=2$ subpopulations}
In this section the equilibrium solutions and stability of the equilibria are discussed for the system (\ref{eq.2}) with the mechanism of species turnover. In Fig.~\ref{fig6} the regions of stable (unstable) equilibria are plotted versus the enrichment and the transfer rate. The results are contrasted on the stability diagrams in Figs.~\ref{fig6} $a-d$
for different predation rates of the IG predators.
Four different states are localized in the parameter space that corresponds to stable (unstable) persistence and exclusion of the IG prey (IG predator). We investigate how the dynamics in the extended IGP system responded to variation of enrichment levels.
For each case shown in  Fig.~\ref{fig6} an increase of enrichment is accompanied with a series of bifurcations in the system manifested by an invasion of higher trophic levels similar to predictions from the linear food chain theory \citep{Oksansen1981}. For instance, at low enrichment both regimes $4$ and $1$ are stable. Further increase of $K$ at fixed $t_r=0.05$ results in a chain of bifurcations from a stable regime $4$ to an unstable $2$ and subsequently to a stable coexistence regime $1$. A further increase of carrying capacity favours an exclusion of $G$ and shifts the population densities towards the dominance of the IG predator. An interesting feature is that at low transfer rate only a coexistence of the IG predator and the resource is found. The second subpopulation $S_2$ is extinguished fast due to predation and low transfer rate. The steady states found for low enrichment are similar to the case of a single prey population without transfer mechanism at $t_r=0$ and $S_2=0$. As typified on the diagram in Fig.~\ref{fig6} d the IG prey levels remain positive. Since the IG prey has an advantage as a competitor for the shared resource only the IG predator gets excluded from the system.

The stability behaviour of the system (\ref{eq.3}) is highly sensitive to the alternations of attack rates of $G$ and the productivity of resource $r$. Changes of these parameters produce different emergent patterns as shown in Fig.~\ref{fig7}.
 The location of states of stable (unstable) permanence and the exclusion zone of $G$ is still comparable to the patterns shown in Fig.~\ref{fig6}, however the region of 3-species coexistence gets visibly reduced. The reduction is more evident on the plots Fig.~\ref{fig7} $a,c$ and $d$. At higher transfer rates the coexistence of all 3-species is no longer observed and only the population of IG prey and resource persist.  Due to low productivity the densities of the basal resource $S_1$ are quickly depleted and the IG predator is driven to extinction. On the contrary, conditions become more profitable for the IG prey that is released from the IGP pressure and simultaneously obtains more benefits by predation on the extra resource $S_2$.

At low transfer rates (Fig.~\ref{fig7}c and d) the IG predator is excluded independently on carrying capacity of the resource. As expected, with increase of the attack rate of $G$ the population of the IG predator is driven to extinction due competition with IG prey. However, situation becomes more favourable for the IG predator at higher values of the transfer coefficient $t_r$. For high enrichment and intermediate transfer the IG prey is excluded from the system. At a fixed enrichment several alternating states are found along the gradient of $t_r$ (see Fig.~\ref{fig7}~d). For example, at $K>2.5$ the behaviour of the food web is very sensitive even to a small alternations of $t_r$. Indeed, the system passes through distinct steady states just within a small increment of transfer rate. The exclusion of the IG predator is observed at $t_r<0.05$, the coexistence is found at $t_r~0.06$ and the exclusion of the IG prey is achieved at $t_r~0.07$. Finally at a higher transfer values ($t_r>0.14$) both predators enter the system and persistence is reached.

After presenting the results for the systems (\ref{eq.2}) and (\ref{eq.3}) we proceed to a more complex situation with $n>2$ of distinct subpopulations of the resource.

\subsubsection{Case of $n>2$ prey subpopulations}

For a multipopulation model the choice of parameters including predation rates can be enormously large. As a consequence more freedom is provided for choosing equilibrium densities that can fit the model (\ref{eq.1}). Since it is impossible to investigate the entire range of biologically plausible parameters we make a particular choice of parameters that allow an easier comparison of the case $n>2$ in (\ref{eq.1}) with the model (\ref{eq.2}). The details of the procedure are provided in Appendix.

In this section we show the results of the numerical simulation for the model with $n>2$ prey subpopulations. The system~(\ref{eq.1}) for the case $n>2$ is integrated numerically. For the calculation of the stability diagrams at different fixed values of enrichment and transfer rate we perform $300$ simulations. The results of the simulations for $n=2,3,4$ and $7$ subpopulations are illustrated in Fig.~\ref{fig8}. The percentage of stable 3-species coexistence
 is calculated for every point in the parameter space with fixed enrichment $K$ and limiting value $t_r$. The colorcode is assigned according to the percentage of stable coexistence solutions found for $300$ food webs. In all the replicas of the simulated system the steady state densities for $G, F$ and $S_1$ are fixed (see (\ref{eq.12}) in the Appendix). Thus only the variations among possible equilibrium densities $\{S_k\}_{k\neq1}$ are examined. The constraints for the parameters of high dimensional system (\ref{eq.1}) are given in eqs.~(\ref{eq.14})-(\ref{eq.16}) in the Appendix.

The stability diagrams in Fig.~\ref{eq.8} show some similarities to the regions of coexistence in ~Figs.~\ref{fig6} and \ref{fig7} found for the
$n=2$ subpopulation model (\ref{eq.3}). The size of the stability zone expands with the increase of the transfer coefficient.  At low transfer rates no stable persistence is found, but different alternative traits. The percentage of stable food webs with 3-species is substantially lower for a large system with $n=7$ subpopulations than for $n=2,3$. This reduction in stability is independent on the number of simulated food webs and a choice of main parameters of the system. It is possible that an increase of food web connectivity in this case impacts negatively the system (\ref{eq.1}) stability. Another feature is that for $n=7$ the percentage of stable equilibria at a fixed enrichment value decreases for large values of $t_r$ unlike in previous cases in Fig.~\ref{fig8} a-c.

The results of the numerical simulation demonstrate that for $n=2$ subpopulation up to $95\%$ of stable systems are found at a higher transfer rate and an intermediate enrichment. Second, for a larger food web with $n=7$ subpopulations a higher percentage of stable steady states (up to $45\%$) are identified at low transfer rate and at high enrichment.

We compare the results of simulation for four cases $(n=2,3,4,7)$ at fixed enrichment $K=1.341$ and variable transfer coefficient in Fig.~\ref{fig9}. The yields are derived for $1000$ simulations of food webs. The estimations of the number of steady states show that for food webs with $n\leq 4$ a higher percentage of solutions with a stable coexistence are identified than for food web with $n=7$ pools. Indeed, the yield for $n=2$ reaches almost $95\%$ meanwhile the percentage of stable food webs found for $n=7$ saturates at $23\%$ for large $t_r$. The non-monotonic variations of the yields in Fig.~\ref{fig9} reveal a highly sensitive behaviour of the IGP model (\ref{eq.1}) to a change in transfer rate in all cases. For $n=7$
the percentage of stable food webs reaches $41\%$ at a low transfer rate. It decreases substantially for higher values of the transfer rate. For $n\leq 4$ there is an overall incline from $60\%$ at $t_r\sim0.1$ to $95\%$ at $t_r\sim0.3$ of stable configurations.

Two types of stable equilibrium solutions are illustrated in Fig.~\ref{fig10}. Both solutions are obtained inside the stable coexistence region as indicated in Fig.~\ref{fig9}. The system (\ref{eq.1}) is simulated with $n=5$ number of subpopulations and initial conditions as defined in the Appendix. For the steady state in Fig.~\ref{fig10}~a and the oscillatory state in Fig.~\ref{fig10}~b most of resource subpopulations are unstable and their densities rapidly decline to zero after some initial transient. Nevertheless, coexistence in the system is typically supported by one or two resource pools with non-zero densities.
\begin{figurehere}
\begin{center}
  \includegraphics[width=9.cm]{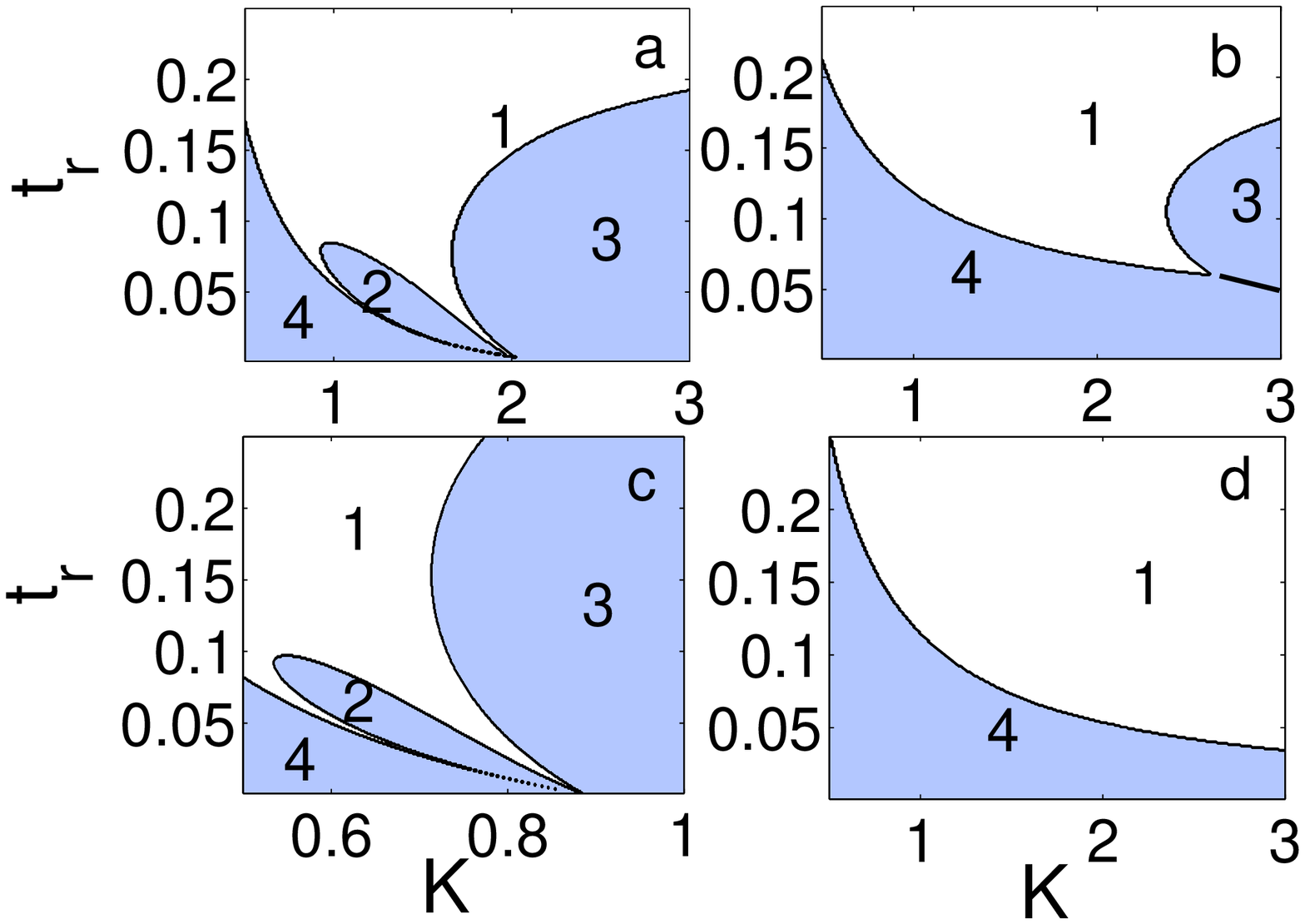}\\
  \end{center}
  \caption{Regions of stable (unstable) equilibria are marked according to species composition:
   $1 (2)$ - stable (unstable) coexistence of resource and both predators $G$ and $F$; $3$ - exclusion of IG prey; $4$ - exclusion of IG predator. The predation rates are: (\textbf{a})~$f=0.1,~a=0.0155$ ;(\textbf{b})~$f=0.01,~a=0.0155$; (\textbf{c})~$f=0.2, ~a=0.0155$; (\textbf{d})~$f=0.1,~a=0.065$. The remaining parameters are: $r=0.5,b=0.1,a'=0.8,g=0.07,f'=0.2,g'=0.5,m_g=0.02,m_f=0.04$}\label{fig6}
\end{figurehere}
\begin{figurehere}
\begin{center}
  \includegraphics[width=9.cm]{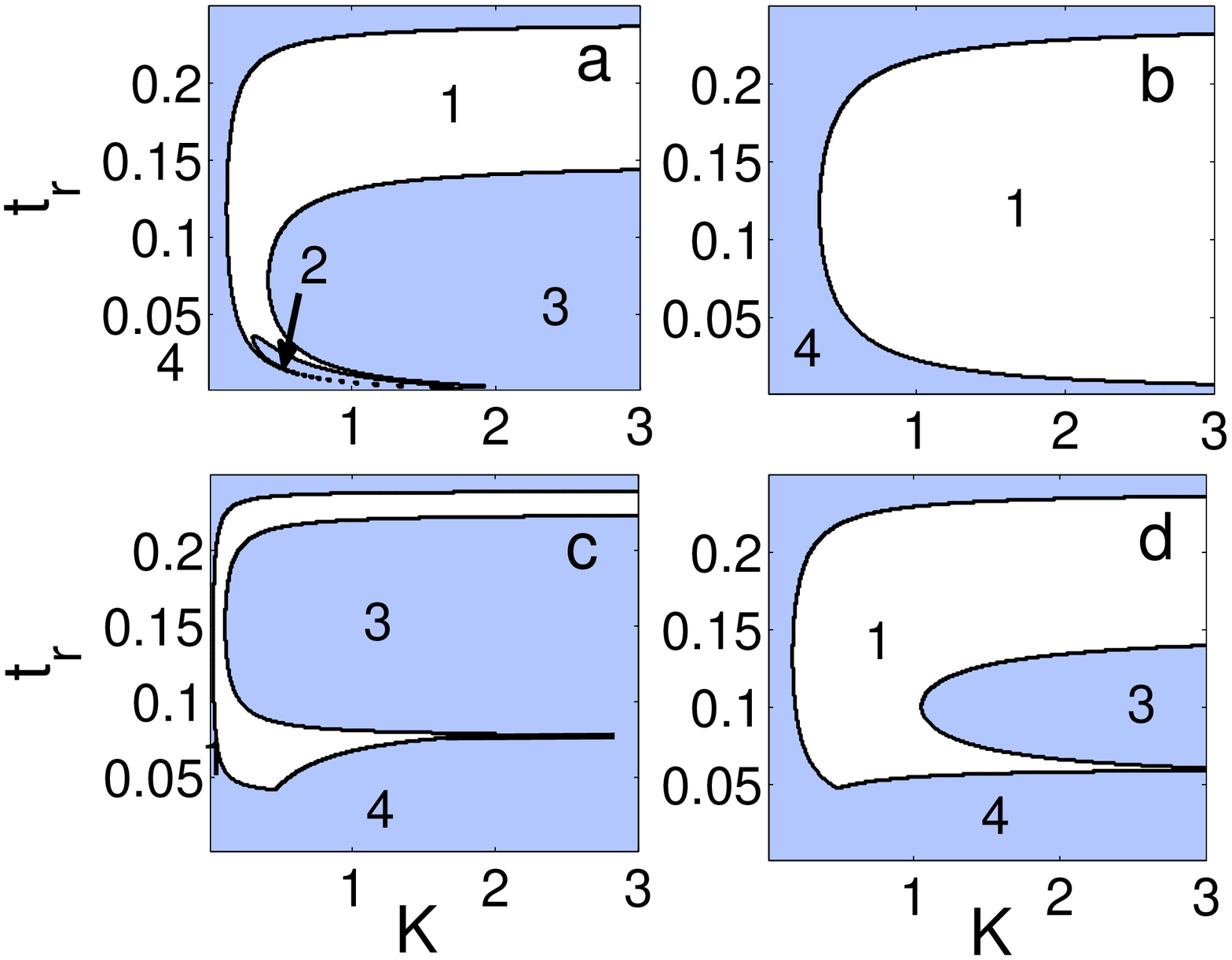}\\
  \end{center}
  \caption{Regions of stable (unstable) equilibria are marked according to stable trophic configurations:
   $1 (2)$ - stable (unstable) coexistence; $3$ - exclusion of IG prey; $(4)$ exclusion of IG predator. The predation rates are: (\textbf{a})~$f=0.1,~a=0.0155$ ;(\textbf{b})~$f=0.01,~a=0.0155$; (\textbf{c})~$f=0.15, ~a=0.065$; (\textbf{d})~$f=0.1,~a=0.065$. The remaining parameters are: $r=0.3,b=0.02,a'=0.8,g=0.07,f'=0.2,g'=0.5,m_g=0.02,m_f=0.04$}\label{fig7}
\end{figurehere}

\begin{figurehere}
  \begin{center}
  \includegraphics[width=9.cm]{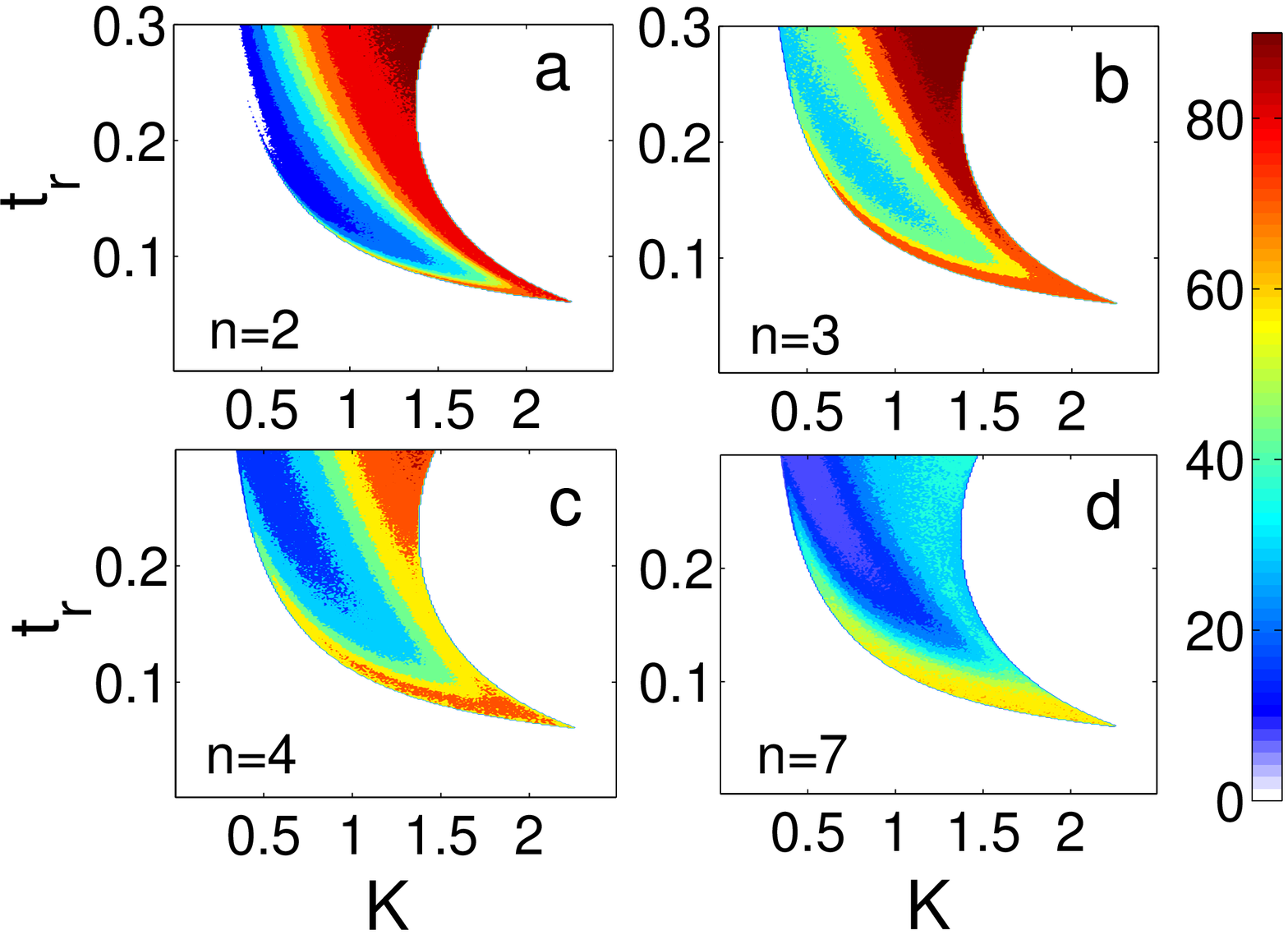}
  \end{center}
  \caption{The percentage of 3-species stable coexistence found for the model~(\ref{eq.1}). The stability region is presented versus enrichment and transfer rate. Simulations for the four cases are given: (\textbf{a}) $n=2$, (\textbf{b}) $n=3$, (\textbf{c}) $n=4$, (\textbf{d}) $n=7$.  Parameters are: $r=1,f=0.1,b=0.1,a'=0.8,g=0.07,f'=0.2,g'=0.5,m_g=0.02,m_f=0.04$}\label{fig8}
\end{figurehere}
\begin{figurehere}
  \begin{center}
  \includegraphics[width=8.cm]{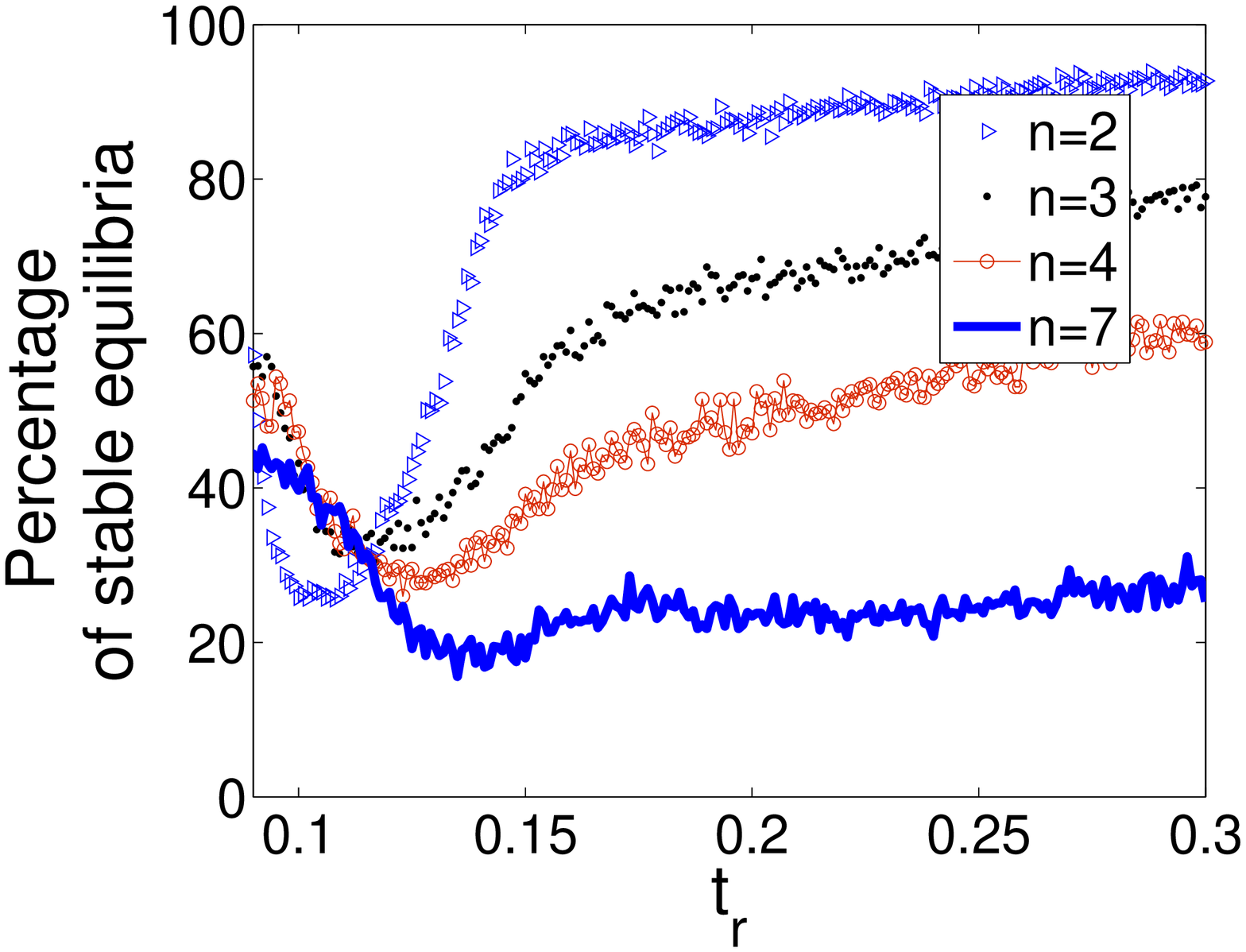}
  \end{center}
  \caption{The percentage of 3-species stable coexistence for the model~(\ref{eq.1}). $N_{sim}=1000$ simulations are performed for the cases: $n=2,3,4,7$. Parameters are as in Fig.~\ref{fig8}. Enrichment: $K=1.341$.}\label{fig9}
\end{figurehere}

\begin{figurehere}
\begin{center}
  \includegraphics[width=9.cm]{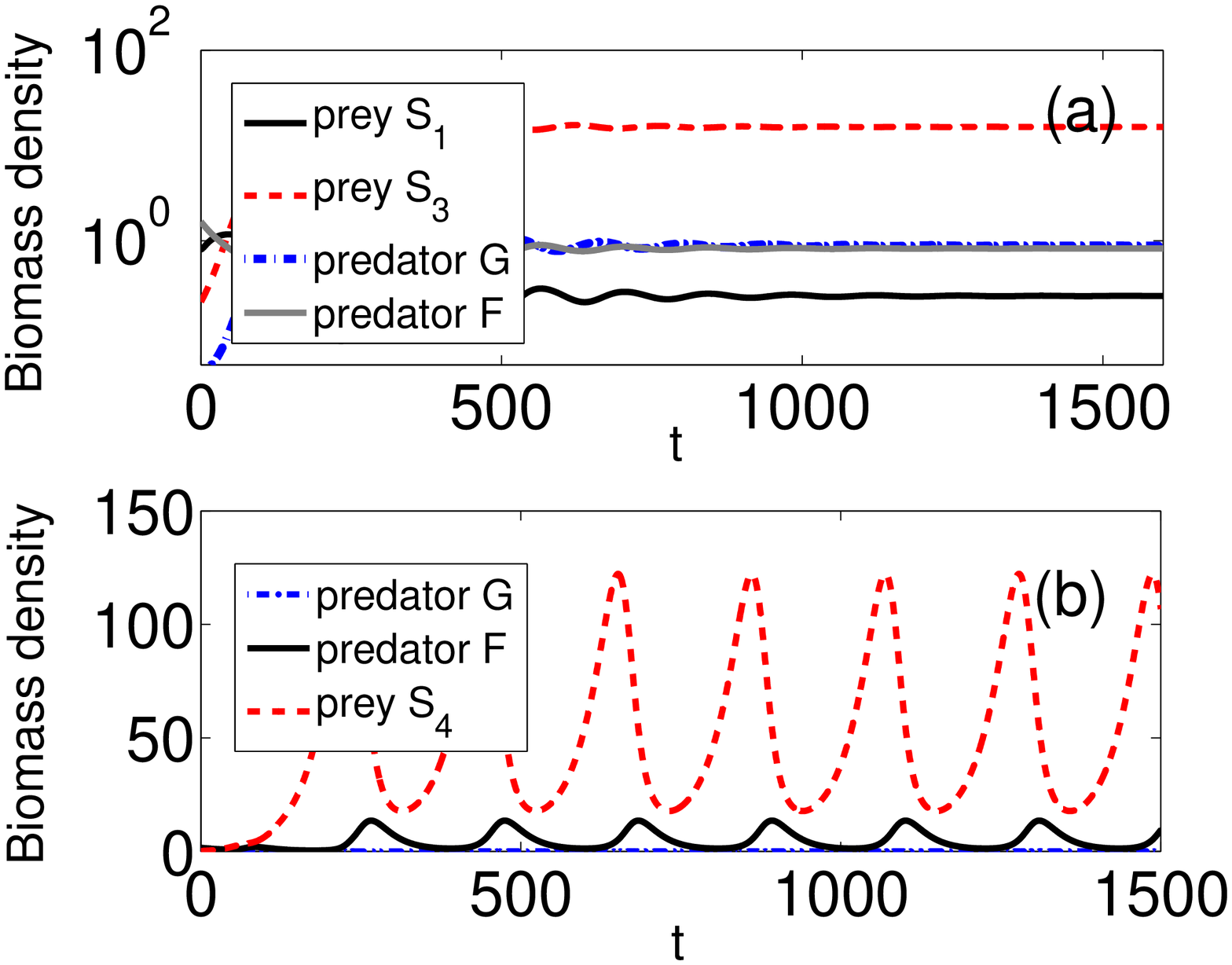}\\
  \end{center}
  \caption{Two stable solutions for the system (\ref{eq.1}) with $n=5$ subpopulations. The equilibrium densities and interaction rates are  described in Appendix. Parameters are: $r=0.7,f=0.1,b=0.1,a'=0.8,g=0.07,f'=0.2,g'=0.5,m_g=0.02,m_f=0.04, K=1.8, t_r=0.2$}\label{fig10}
\end{figurehere}
\section{Discussion}\label{sec.5}

There is growing evidence from theoretical and empirical studies that creating additional trophic links have a stabilizing effect on food webs \citep{Moore2005,Ives2007}. Generalized models reveal that the stability of food webs can be enhanced when
 species at higher trophic levels graze upon multiple prey species~\citep{gross2009}. In particular, for low dimensional food webs it is demonstrated that an addition of alternative food resources can stabilize the interactions \citep{Holt2007} and open up a possibility for feedbacks on population dynamics due to apparent competition. The predictions of our model confirm the main conclusions given in a theoretical study of an extended IGP model \citep{Holt2007}. In the alternative formulations used here the IG prey has the access to an extra  resource beyond the shared resource for which both predators compete. This extra resource is a more attractive resource item for the IG prey and is thus attacked at higher rates by the IG prey whereas the attack rate of the IG predator stays the same.  Moreover the IG predator indirectly stimulated the growth of the IG prey population by providing this extra resource. Our predictions tested by the application of a stability analysis are robust in the sense that they are independent of the form of the interaction term that is responsible for the availability of an additional resource.  We demonstrate that for different formulations of the basic IGP model with the embedded interactions a stable 3-species coexistence is ultimately reached whenever a moderate strength of the omnivorous links is used.

However, the problem to relate the experimental findings \citep{Loeder2012} to the theoretical predictions of our model (\ref{eq.2}) still remains open. Since the experiment is aimed to observe a short term populations development it is not easy to find a direct correspondence between empirical population dynamics and theoretically predicted behaviour. In the $3$ day batch culture experiment of L{\"o}der et al. with all 3 species present both predators {\it G. dominans} and {\it F. ehrenbergii} displayed positive growth while the prey population {\it S. trochoidea} displayed simultaneously a sharp decline to almost zero. How can this behaviour be classified according to our theoretical model? Could it be a part of an oscillatory cycle or an unstable state? It is not easy to answer these questions, however, we can make a guess that the short term evolution observed in the experiment recasts as a part of an oscillatory cycle for a periodic equilibrium state found at intermediate immobilization. A similar type of experiments performed for various initial species densities could furnish a justification of this hypothesis.

The above results demonstrate that a persistence of IG predator, IG prey and resource is achieved even at a low value of immobilization rate. Moreover, a significantly higher percentage of observed stable configurations is found when the immobilization and transfer links in (\ref{eq.1}) and (\ref{eq.2}) are strengthened. Because our model is an oversimplification of the experimental behaviour the partitioning of the parameter space according to stable versus unstable coexistence could be used as an approximation of the population dynamics found in a real experimental situation. Firstly, the conditions for long term stable coexistence found by numerical simulations are not so easy to examine experimentally because of technical and temporal restrictions. Experimental samples in ref. \citep{Loeder2012} are taken during $3$ days of incubation due to a decline of the prey population. Secondly, due to the existence of stable limit cycles  as predicted by our linear stability analysis (see Fig.~\ref{fig10}) the oscillatory solutions  go through a period of very low densities and might be driven to extinction in the presence of random fluctuations of the environment.

We point out that our numerical simulations of the extended 3-species IGP module (\ref{eq.1}) do not explore the entire swath of parameters and configurations for the steady states. Rather our analysis focuses on explaining conceptual features of the IGP model with diverse prey populations.

Earlier studies focussing on dynamics of complex ecological communities  \citep{Dewitt2003} demonstrated the importance of multiple prey traits in mitigating predator selection pressures and altering predator-induced behavioural shifts in natural environments \citep{lima1990,trussell2002}.
Our model can also be adapted to food web communities in which differentiation among prey individuals, namely, variation in individual traits such as fitness and mobility, is a result of heterogeneities in their natural habitat and/or adaptation of the species to the local conditions of the habitat. We show that an existing diversity of resource items traits can significantly alter the emergent community patterns. Adding new subpopulations of resources with distinct traits that are more vulnerable to an attack from the IG predator facilitates the coexistence of both IGP-related predators which compete for the common food resource. Thus, the presence of an alternative resource indirectly induces shifts in exploitative competition.

It is important to note that a general mathematical model with density- dependent interactions and immobilization do not render a unique theoretical description for the results of the experiment \citep{Loeder2012}. Our model predictions can be tested against alternative formulations. Indeed, the main features of the experimental system can be examined by the inclusion of predation rates that are dependent on the mobility of the resource species. Since slow and immobile individuals can also be found among mobile species one can use an inhomogeneous distribution of velocities of the resource species in a theoretical model. To guarantee more benefit for an intermediate consumer in catching a certain type of individual distinct predation rates should be assigned according to different velocities of resource species. Another question is if the growth rate of the IG predator will be affected by the inclusion of the time of resource capture. How will the inclusion of the time lag change the predictions of our immobilization model?
These extensions of a general IGP model will be a topic for our future investigations.

Finally, we point out that it is of potential interest for biological control and conservation management to understand functioning of omnivory and IGP systems in relation to global changes of the environment. Since IGP food webs are widespread in natural communities their adaptation and resilience behaviour is principal for understanding the restructuring of natural communities.

\section{Conclusions}

We have used three formulations of a general IGP model to explore the effects of increasing diversity in the prey population on higher trophic levels. The reformulated IGP model alters the results from the basic IGP theory \citep{polis1992,Holt1997,diehl2000}. We show that an increase of a number of trophic interactions in the system via differentiation of resource can stabilize the population dynamics of the IGP module. This conclusion holds for the densities of the IG prey that level up even when the IG predator is a superior competitor for the common basal resource.

The results of our numerical simulations can be summarized as follows.

First, we show that for the system with the immobilization term up to three regions of stable trophic configurations are observed along the enrichment gradient. Meanwhile at low enrichment both IG prey and IG predator are excluded, at high enrichment the presence of only small concentration of immobilized cells is sufficient to facilitate the coexistence  of the competitors in the IGP relationship. Moreover the percentage of all admissible trophic configurations for the 3-species persistence inclines substantially for higher immobilization.

Second, given that immobilization is high enough it prompts the exchange between pools of mobile and immobilized resource and facilitates fast decline of the mobile population and a growth of immobilized subpopulation. Meanwhile the exchange between the basal mobile resource and the predators gets weaker due to the low density of mobile species the immobilized individuals become a major food resource for the predators. Because the IG prey is a superior competitor for immobilized resource a robust coexistence of both predators will be easily supported. In addition, along an increasing gradient of immobilization the relative abundance of IG prey becomes higher than the abundance of IG predator.

Restructuring of the basic IGP module by adding individual-to-individual turnover facilitates the coexistence and stabilizes the otherwise unstable system. Moreover a strengthening of the interaction link leads to a significantly broader range of enrichment values at which stable coexistence could be found. At low transfer rate two types of equilibria are observed: $(i)$ if the IG predator is a superior competitor for the resources than at low enrichment both predators are excluded and at high enrichment only the IG predator stays in the system; $(ii)$ an increase of the attack rates of the IG prey depresses the population of the IG predator until it is completely excluded.

Numerical simulations of food web (\ref{eq.1}) with $n=2,3,4$ and $7$ distinct pools demonstrate that the high dimensional food webs overall manifest far less stable behaviour than the food webs with only two distinct subpopulations. An interesting feature is that the percentage of stable states for $n=7$ substantially decreases from $40\%$ to $23\%$ with an apparent increase of the value of transfer rate. By contrast, for food webs with $n\leq 4$ an increase in transfer rate leads to the growth of the percentage of stable coexistence solutions from about $60\%$ to $95\%$.

\appendix
\section{}\label{sec.6}
In the Appendix we review the steady state solutions for the Lotka-Volterra models (\ref{eq.2}) and (\ref{eq.3}) and provide Jacobian matrices to examine their local stability for the coexistence of both predators and the resource.

First, equilibrium solutions are derived for
the 3-species model~(\ref{eq.2}) with zero immobilization ($i_m=0$) and zero initial size of immobilized population ($S_i=0$).
Second, the steady states are given for the model ~(\ref{eq.2}) with immobilization ($i_m\neq 0$). At last, the equilibrium solutions are presented for the system with the resource turnover (\ref{eq.3}).
For every case various trophic configurations are considered: $(i)$ exclusion of both predators, $(ii)$ exclusion of IG prey or IG predator and  $(iii)$ the 3-species coexistence.

\subsection{Steady state solutions for 3-species model}
The equilibrium solution of~(\ref{eq.2}) for the 3-species coexistence without immobilization
is stated as follows:
\begin{eqnarray}
S_{eq}&=&\bigg(r-\frac{a m_f}{g g'}+\frac{m_g f}{g}\bigg )\nonumber\\
&&\times\bigg(\frac{r}{K}-\frac{f a}{g}\bigg( \frac{f'}{g'}-a'\bigg)\bigg)^{-1},\nonumber\\
G_{eq}&=&(m_f-f f' S_{eq})(g'g)^{-1},\nonumber\\
F_{eq}&=&(a'a S_{eq}-m_g)g^{-1},\label{eq.4}
\end{eqnarray}
where $g, g'>0$. The necessary condition for the coexistence requires that the right hand side is positive in (\ref{eq.4}).

The expressions for the equilibrium densities for the survival of the IG prey and the resource with exclusion of the IG predator at $F_{eq}=0$ yield:
\begin{equation}
S^F_{eq}=\frac{m_g}{a a'},~
G^F_{eq}=\frac{r}{a}\bigg(1-\frac{S^F_{eq}}{K}\bigg).\label{eq.5}
\end{equation}
 The condition for persistence of the IG prey and the resource reads: $a'a K>m_g$.

At zero density of the intermediate consumer ($G_{eq}=0$) one yields the steady states of the resource $S^G_{eq}$ and the IG predator $F^G_{eq}$:
\begin{equation}
S^G_{eq}=\frac{m_f}{f f'},~
F^G_{eq}=\frac{r}{f}\bigg(1-\frac{S^G_{eq}}{K}\bigg).\label{eq.6}
\end{equation}
The densities are positive if and only if the condition $f f' K>m_f$ holds true.
\subsection{Steady state solutions and linear stability analysis}
\subsubsection{Model with immobilization}
Here we describe alternative steady state solutions and discuss their local stability derivation. Also the Jacobian matrix for the 3-species coexistence is provided in the explicit form.

For the model~(\ref{eq.2}) with immobilization $(i_m\neq0)$ we define a set of equilibrium densities to satisfy the equalities below:
\begin{eqnarray}
Q_1&=&[r(1-S_m K^{-1})-a G\nonumber\\&&-(f+i_m)F]S_m,\nonumber\\
Q_2&=&i_m F S_m-S_i( b G+f F ),\\
Q_3&=&(a'a S_m+a'b S_i-g F-m_g )G,\nonumber\\
Q_4&=&(f'f( S_i+S_m )+g g' G-m_f )F.\nonumber\label{eq.7}
\end{eqnarray}
 The system (\ref{eq.7}) has four alternative solutions : $(i)$ exclusion of both predators at  $S=K$; $(ii)$ exclusion of the IG predator $(S_m^F,G^F)$ at $F=0$; $(iii)$ the coexistence of resource and the IG predator $(S_m^G,S_i^G,F^G)$ at $G=0$; $(iv)$ the 3-species coexistence $(S_m^e,S_i^e,G^e,F^e)$.

In the absence of $F$ the immobilization mechanism is not active and the model (\ref{eq.2}) reduces to the system without immobilization~\citep{diehl2000}
where the equilibrium solutions written as (\ref{eq.5}).
Upon exclusion of the IG prey in~(\ref{eq.7}) one obtains expression for the equilibrium densities of resource and the IG predator:
\begin{eqnarray}
S^G_m&=&\frac{m_f}{ f'(i_m+f)},~
S^G_i=\frac{i_m}{f}S^G_m,\nonumber\\
F^G&=&\frac{r}{f+i_m}\bigg(1-\frac{S^G_m}{K}\bigg).\label{eq.8}
\end{eqnarray}
Note that the size of mobile population $S^G_m$ is proportional to the size of immobilized population $S^G_i$.
As is expected the population of immobilized preys is impacted positively by the increase of immobilization. Although the predation pressures are equal for both resource subpopulations the immobile population $S^G_i$ extinguishes faster than the mobile population $S^G_m$. Indeed, with the increase of predation rate $f$ the following approximations hold: $S^G_m\sim 1/(i_m+f)$ and $S^G_i \sim 1/f(i_m+f)$.

The equilibrium densities for the 3-species coexistence are derived from ~(\ref{eq.7}) by setting the right hand side to zero.

At last, to evaluate the stability of the equilibrium solution $S^e_m, S^e_i ,G^e$ and $F^e$ one solves for the eigenvalues of the stability matrix :
\vspace{5 mm}

$ \left[ \begin{array}{cccc}
      D_1 & 0 & -a S^{e}_m & -(f+i_m) S^{e}_m \\
      i_m F^{e} & D_2 & -b S^{e}_{i} & i_m S^{e}_m-f S^{e}_i \\
      a a' G^{e} & b a' G^{e} & D_3& -g G^{e} \\
      f f' F^{e} & f f' F^{e} & g g' F^{e} & D_4 \\
    \end{array}\right]$
\vspace{5 mm}

The matrix diagonal is written in terms of the equilibrium densities $S^e_m, S^e_i,G^e$ and $F^e$ as follows:
\begin{equation}
 D=\bigg(-\frac{r}{K} S^{e}_m,-b G^{e}-fF^e,0,0\bigg).\label{eq.9}
 \end{equation}
 The solution ($S^e_m,S^e_i,G^e,F^e$) is globally asymptotically stable in the phase space \citep{Svirezhev1983} if the condition for stability is satisfied.
For the stable coexistence it is necessary that the real parts of all four eigenvalues $\lambda_{i},(i=1,\ldots 4)$ of the stability matrix  are non-positive. To obtain the boundary for stability regions in the parameter space the eigenvalues of the above stability matrix are evaluated numerically at different parameters combinations. The resulting stability diagrams are presented in Fig.~\ref{fig2} and Fig.~\ref{fig5}.

\subsection{Model with prey--to--prey interactions and $n=2$ subpopulations}

As in the previous case the system \ref{eq.3} for $n=2$ pools permits four steady states:~$(i)$ the exclusion of the predators at $S_1+S_2=K$; $(ii)$ the exclusion of the IG predator $(S^F_1,S^F_2,G^F)$ at $F=0$; $(iii)$ the exclusion of the IG prey $(S^G_1,S^G_2,F^G)$  at $G=0$; $(iv)$ the coexistence of 3-species $(S_1^e,S_2^e,G^e,F^e)$.

The solution for the coexistence of the resource and the IG prey in the absence of the IG predator is expressed as follows:
\begin{eqnarray}
S^F_1&=&K\bigg(1-\frac{t_r m_g}{r a'b}\bigg),\nonumber\\
S^F_2&=&\frac{m_g-a'a S^F_1}{a'b},~G^F=\frac{t_r}{ b} S^F_1.\label{eq.10}
\end{eqnarray}
Note that at $t_r=0$ the IG prey is excluded and steady state density for the resource approach the carrying capacity limit $K$. The positive solution of (\ref{eq.10}) exists if the parameters satisfy the inequality: $r a' b >t_r m_g$.

The steady state for the resource and the IG predator in the absence of IG prey yields:
\begin{eqnarray}
S^G_1&=&K\bigg(1-\frac{t_r m_f}{f'f r} \bigg),\nonumber\\
S^G_2&=&(1+K t_r)\frac{m_f}{f'f}-K,\nonumber\\
F^G&=&\frac{K t_r} {r f}\bigg(r-t_r\frac{m_f}{f'f}\bigg).\label{eq.11}
\end{eqnarray}
 A nontrivial solution for the 3-species coexistence is found by solving for the equilibrium $(S^e_1,S^e_2,G^e, F^e)$ of the following system:
\begin{eqnarray}
0&=&r(1-S^e_1 K^{-1})-a G^e\nonumber\\&&-f F^e-t_r S^e_2,\nonumber\\
0&=&t_r S^e_1-b G^e-f F^e,\nonumber\\
0&=&a'a S^e_1+a'b S^e_2-g F^e-m_g,\nonumber\\
0&=&f'f (S^e_1+S^e_2)+g'g G^e-m_f.\label{eq.12}
\end{eqnarray}
Finally, the condition for the stable coexistence is provided by solving for the eigenvalues of the stability matrix :
 \vspace{5 mm}

 $ \left[ \begin{array}{cccc}
      -r /K S^{e}_1 & -t_r S^e_1 & -a S^e_1 & -f S^e_1 \\
      t_r S^e_2 & 0 & -b S^e_2 & -f S^e_2 \\
      a a' G^e & b a' G^e & 0 & -g G^e \\
      f f' F^e & f f' F^e & g g' F^e & 0 \\
  \end{array}\right]$\nonumber \\
The eigenvalues of the stability matrix are functions of the constant rates (see Table.~\ref{tab2}) and the equilibrium densities $S^e_1,S^e_2,G^e$ and  $F^e$. The condition for the stable coexistence requires that the real parts of the eigenvalues of the stability matrix are non-positive.
\subsection{Model with prey--to--prey interactions and $n>2$ subpopulations}

Here we define initial conditions that are used for the numerical simulation of the model (\ref{eq.1}) with $n>2$ number of subpopulations. 
 Unlike for the system (\ref{eq.3}) with $n=2$ subpopulations the equilibrium densities can no longer be determined analytically. To simplify the search for the equilibria in (\ref{eq.1}) we implement several assumptions.

We chose the parameters and equilibrium densities to fulfil several constraints provided below. Initial densities are equal to the steady states (\ref{eq.12}) for $n=2$ subpopulations, namely: $S_1=S^e_1,S_2=S^e_2, G=G^e$ and $F=F^e$. For the sake of simplicity the values for the remaining densities $\{S_k\}^{n-1}_{k=2}$ are defined to be less than the equilibrium density $S^e_2$. Provided that the interaction rates $\{c_k\}_{k\neq 1}$ are randomly assigned values not exceeding $t_r/(n-1)$ the equilibrium density $S_n$ can be found from the constraint:
\begin{eqnarray}
\sum^n_{j=1}c_j S_j=t_r S^e_2.\label{eq.14}
\end{eqnarray}
 We chose the zero rates of decline $m_1$ and $m_2$ for the first two subpopulations $S_1$ and $S_2$. The values for the grazing rates  $\{f_k\}_{k>3,}$ are randomly assigned from the interval $[0, f/n]$. In the next step an antisymmetric matrix $Q$ of individual-to-individual  interactions is defined with the upper diagonal coefficients $\{q_{j k}\}_{j<k}$ that obey the inequalities:  $-0.1~t_r\leq q_{j,k}\leq 0.1~t_r$.
The remaining lower diagonal coefficients $\{q_{k j}\}_{j<k}$ should satisfy  the antisymmetry relation: $q_{j_k}=-q_{k j}$. The set of equations in~(\ref{eq.1}) for $k=3,\ldots n$ holds true if the positive mortality rates $\{m_k\}_{k>2}$ are expressed from the equations as follows:
\begin{eqnarray}
m_k=c_k S_1-\sum q_{j k} S_j-b_k G-f_k F.\label{eq.15}
\end{eqnarray}
The equation (\ref{eq.15}) for $k=2$ is used to solve for the attack rate $f_2$.
Finally, the attack rates $\{b'_k\}^{n-1}_{k=2}$ and the feeding rates $\{f'_k\}^{n-1}_{k=2}$ are randomly
 assigned from the intervals $[0, a'b/(n-1)]$ and $[0, f'f/(n-1)]$ correspondingly. The choice of the feeding rates enables to reduce the predation pressure on populations $\{S_k\}^{n-1}_{k=2}$ by a factor of $1/(n-1)$. The remaining grazing rates $f'_1$ and $b'_1$ are defined from the relations: $f'_1=f'f$ and $b'_1=a' a$. Finally, two constraints hold to solve for the coefficients $b'_n$ and $f'_n$:
\begin{eqnarray}
\sum^{n}_{j=2}f'_j S_j=f'f S^e_2,~\sum^n_{j=2} b'_j S_j=a'b S^e_2.\label{eq.16}
\end{eqnarray}
We want to emphasize that the attack rate $b'_1$ should be higher than any of the rates $\{b'_j\}^{n}_{j=2}$.  Nevertheless since $a>b$ relation holds the total predation pressure of $G$ summed over alternative pools exceeds the predation exclusively on $S_1$. This far the positive   population density of the IG prey is maintained due to the consumption of alternative resources $\{S_j\}_{j\neq2}$.

We provided a special assignment of parameters for the predation, feeding and mortality that enables to fulfil the condition of positive equilibria that are comparable to the realistic biodensities. Moreover the stability results can be conclusively compared with two different formulations (\ref{eq.1}) and (\ref{eq.3}).

\end{multicols}
\end{document}